\documentstyle[aps,twocolumn]{revtex}
\begin{document}
\title{The General Structure of Bose Gases with Arbitrary Spin}
\author{Tin-Lun Ho and Lan Yin}
\address{Department of Physics,  The Ohio State
University, Columbus, Ohio 43210}

\maketitle

\begin{abstract}
Motivated by the recent discoveries of spin-1 and spin-1/2 Bose gas, we have 
studied the general structure of the Bose gases with arbitrary spin. 
A general method is developed to uncover the
elementary building blocks of the angular momentum eigenstates, as well as 
the relations (or interactions) between them. Applications of this method to 
Bose gas with integer spins $(f=1,2,3)$ and half integer spins $(f=1/2, 3/2)$ 
reveal many surprising structures. 
\end{abstract}

Recent experiments on dilute quantum gases of alkali atoms have produced a
spin-1 and (pseudo-) spin-1/2 Bose gas respectively. The former is produced  in
optically trapped $^{23}$Na\cite{MIT}, the latter     in magnetically 
trapped $^{87}$Rb by rotating the hyperfine states $|f=2, m=1\rangle$ and
$|f=1, m=-1\rangle$ into each other through a 
slightly detuned rf-field\cite{JILA}. Many macroscopic 
quantum phenomena have been observed in these systems. 
At present, these phenomena can be explained in terms of the single 
condensate picture. However, in the case of spin-1 Bose gas with 
antiferromagnetic interaction like $^{23}$Na, it is been pointed out very 
recently 
that as the magnetic field gradient is reduced, the single condensate will
evolve toward an angular momentum eigenstate, which will become 
a spin singlet as the magnetic field is reduced to 
zero\cite{Bigelow}\cite{Hoyip}. The singlet state is a ``fragmented" structure
which bears no resemblance to the single condensate state\cite{Hoyip}. 
That the ground state of a Bose system can be very different from a 
conventional single condensate when it acquires internal degrees of 
freedom is a surprise. 

Motivated by the fragmented structure of the spin-1 Bose gas, 
we consider Bose gases with higher spins. 
Although Bose gases with spin
$f>1$ have not yet been produced, it is conceivable that they can be realized
in the future. After all, both spin-1 and spin-1/2 Bose gases have only 
come into 
existence within the last one and half year. Currently, there are also 
efforts to condense $^{85}$Rb, which will be a spin-2 Bose gas when
loaded into an optical trap. The main reasons for our investigation, however, 
remains theoretical and conceptual. The nature of the ground states of 
Bose gases with internal degrees of freedom is of fundamental importance. 
It has a place in the lore of superfluid physics and significance that 
goes beyond to the study of Bose-Einstein condensation. Our goal is to 
present a general method to construct the (total) angular momentum 
eigenstates $|F, F_{z}\rangle$ for Bose gases with arbitrary spin $f$. 
The construction of these
eigenstates is a crucial step in diagonalizing the Hamiltonian of the system. 
Our method reveals many surprising structures. 
Generally, the spin state $|F, F_{z}=F\rangle$
is made up of singlet and magnetic building units.
A schematic representation of the structure of the spin state
$|F,F\rangle$ for Bosons with spins $f=1,2,3$ and ``pseudo-spin"  $3/2$
are shown in figure a to d. They illustrate the intricate 
structure of
these eigenstates and their increasing complexity with increasing $f$.

The essence of the problem can be illustrated by considering  
a homogeneous Bose gas with spin-$f$. (Its relation to 
a trapped gas can be understood either in terms of local density
approximation and in the procedure outlined in ref.\cite{Hoyip}.)
For a homogeneous dilute Bose gas, we first consider 
the condensate in the zero 
momentum mode (i.e. ${\bf k=0}$), denoted by the annihilation operator 
$a_{\mu} \equiv a_{\mu}({\bf k}=0)$, where $\mu$ labels the $2f+1$ spin 
components. The angular momentum operator then becomes ${\bf \hat{F}}=
a^{\dagger}_{\mu}{\bf f}_{\mu \nu}a^{}_{\nu}$,  where ${\bf f}_{\mu \nu}$ 
is the spin matrix for a spin-$f$ Boson. The effect of the ${\bf k}\neq 0$ 
modes is to deplete the condensate. However, they can be ignored in the zeroth
order approximation as they only contribute a small correction to the energy. 
(For trapped gases, the ${\bf k}=0$ mode will be  replaced by the lowest
self-consistent mode that the system condenses
into\cite{Bigelow}\cite{Hoyip}). To construct the angular momentum
eigenstates,  it is sufficient to focus on the states 
$|F, F_{z}=F\rangle$ with maximum spin projections, since other states with 
$F_{z}<F$ can be obtained by  applying to $|F,F \rangle$
the spin lowering operator 
$\hat{F}_{-}= \hat{F}_{x}-i\hat{F}_{y}$. In the following, 
we shall first derive our method, and then illustrate its application for the
integer cases  $f=1$ to 3 and half integer case $f=1/2$ and 3/2. The case 
$f=3$ is particularly subtle and will be considered last. 

{\em (I.1) Outline of the Generating Function Method}:
To give an orientation of our discussions, we first outline the logic of
our method before presenting the detailed derivations.  We begin by 
considering the total number of  maximum spin states $|F,
F\rangle$ for a systems of
$N$ particles, which we denote as $M_{N}(F)$. To generate this number for all 
$N$ and $S$ simultaneously, we consider the generating function
\begin{equation}
G(x,y) = \sum_{N\geq 0} \sum_{F\geq 0} M_{N}(F) x^{N}y^{F}.
\label{gen} \end{equation}
where $x$ and $y$ are complex numbers within the unit circle ($|x|, |y| <
1$) to ensure convergence. Once this function is constructed, we shall see that
$M_{N}(F)$ is given by the number of solutions of a set of equations obeyed by 
two sets of non-negative integers  $\{ s_{i}\geq 0 \}$ and $\{ m_{j}\geq 0
\}$. The integer $s_{i}$ is the number of singlet building unit 
$\Theta_{i}$ which is made up of $n^{(s)}_{i}$ Bosons and carries no spin,
while   $m_{j}$ is the number of magnetic building unit $\Gamma_{j}$ 
which is made up of $n^{(m)}_{j}$ Bosons and carries spin $\ell_{j}$. 
The integers  $\{ s_{i}\geq 0 \}$ and $\{ m_{j}\geq 0 \}$ satisfy number
and spin constraints  
\begin{equation}
\sum_{i} n^{(s)}_{i}s^{}_{i} + \sum_{j} n^{(m)}_{j} m_{j} = N, \,\,\,\,\,\,
\sum_{j} \ell_{j} m_{j} = F  , 
\label{numberspin} \end{equation}
as well as a set of conditions ${\cal L}_{\alpha}$ that further limit
the range of the integers $\{ s_{i}\}$ and the $\{ m_{j}\}$. These
conditions ${\cal L}_{\alpha}$ reflect the inter-dependence (or
``interactions") among the building units. The conditions ${\cal L}_{\alpha}$ 
are very simple for spin $f<3$ but become quite complicated as 
$f\geq 3$, illustrating the rapidly increasing complexity of the system 
as $f$ increases. 
The typical form of these conditions will become clear when we come to our 
examples. The general structure of
the  maximum spin state is therefore 
$|F, F\rangle = \sum A(\{ s_{i}\}, \{ m_{j}\}) \prod_{i, j} 
\Theta^{\dagger s_{i}}_{i}\Gamma_{j}^{\dagger m_{j}}|{\rm vac}\rangle$, 
where the $A$'s are coefficients and the sum is over all non-negative 
integers $\{ s_{i}\geq 0 \}$ and $\{ m_{j}\geq 0 \}$ satisfying the constraints 
${\cal L}_{\alpha}$. 

{\em (I.2) Derivation of the Generating  Function Method :} 
We begin with the observation that the integer
$M_{N}(F)$ can be expressed as 
\begin{equation}
M_{N}(F) = I_{N}(F) - I_{N}(F+1), 
\label{Mdef} \end{equation}
where $I_{N}(F)$ is the total number of states with 
$F_{z}=F$, independent of the value of total spin $F$. Eq.(\ref{Mdef}) 
follows from the fact that all spin multiplets with total spin $F'>F$ will
contain a state $|F', F_{z}=F\rangle$,  which contribute 1 to both
$I_{N}(F)$ and $I_{N}(F+1)$, and hence $0$ to $M_{N}(F)$. 
 Only those spin states with total spin $F^{\rm
total}=F_{z}=F$ will be included in the integer $I_{N}(F)$ and not
$I_{N}(F+1)$.  That eq.(\ref{Mdef}) is useful is because it is much easier to 
construct a  generating function for 
$I_{N}(F)$ due to the removal of the spin constraint.  Before proceeding,
we note that while 
$M_{N}(F)$ is defined only for $F\geq 0$, $I_{N}(F)$ is defined 
for both positive and negative $F$ such that $I_{N}(F)= I_{N}(-F)$. 

To find $I_{N}(F)$, we note that a 
many-body state with total spin projection $F_{z}=F$ is of the form
\begin{equation}
|F,F_{z}=F\rangle = \sum_{ \{n_{j}\geq 0 \} } 
B(\{ n_{j} \}) \left(\prod_{j=-f}^{f} a^{\dagger n_{j}}_{j}\right) 
|{\rm vac}\rangle
\end{equation}
with $\sum_{j=-f}^{f} n_{j} = N$ and $\sum_{j=-f}^{f} j n_{j} = F$, where 
$\{ n_{j} \}$ is a set of $2f + 1$ non-negative integers, $a^{\dagger}_{j}$
creates a Boson in spin state $j$, and $B$'s are coefficients.  The number of
states with $F_{z}=F$ is 
\begin{equation}
I_{N}(F) = \sum_{\{ n_{j}\geq 0 \} } \Delta\left( \sum_{j=-f}^{f}
n_{j}\right) \Delta\left( \sum_{j=-f}^{f} j n_{j}- F\right)
\label{Idef} \end{equation}
where $\Delta(x)$ is a delta-function ensuring the vanishing of 
$x$.  The generating function eq.(\ref{gen}) can in principle be 
obtained by substituting eqs.(\ref{Mdef}) and (\ref{Idef}) 
into eq.(\ref{gen}). However, the constraint $F \geq 0$ in eq.(\ref{gen}) 
prevents an efficient summation. We therefore consider the function
\begin{equation}
W(x,y) = \sum_{N \geq 0}\sum_{F} \left( I_{N}(F) - I_{N}(F+1) \right) x^{N}
y^{F}, 
\label{Wdef} \end{equation}
where the sum $F$ ranges over all integers. Clearly, $G(x,y)$ is 
$W(x,y)$ with all negative powers of $y$ eliminated. This elimination
can be achieved by the following integration 
\begin{equation}
G(x,y) =  \int_{0}^{2\pi} \frac{{\rm d}\theta}{2\pi}
\sum_{\ell = 0}^{\infty} \left[ (y z^{-1})^{\ell}  W(x,z)  
\right]_{z = e^{i\theta}} \,\,\,\,\, . 
\label{projection} \end{equation}
Performing the sum in eq.(\ref{projection}), $G$ becomes a contour integral 
around the unit circle $C$, $z=e^{i\theta}$,
\begin{equation}
G(x,y) = \int_{C} \frac{ {\rm d}z}{2\pi i}  \frac{ W(x,z)}{z-y}   .
\label{con} \end{equation}
The expression $W$ can be obtained easily since $F$ now runs through all
integers.  Substituting eq.(\ref{Idef}) into eq.(\ref{Wdef}), and first sum
over $F$ and $N$, the functions $W$ becomes
\begin{eqnarray}
 & W(x,z) & = \left( 1 - z^{-1}\right) \left[ \prod_{j=-f}^{f}
\sum_{n_{j}=0}^{\infty} \right] 
 \left( \prod_{j=-f}^{f} x^{n_{j}} z^{jn_{j}} \right)  \\
 &  &= (1 - z^{-1})\prod_{j= -f}^{f} \frac{1}{1 - xz^{j}} , 
\label{Wfinal} \end{eqnarray}
We then arrive at the key expression for the generating function,
\begin{equation}
G(x,y) = \int_{C} \frac{ {\rm d}z}{2\pi i} \frac{1-z^{-1}}{z-y}
\prod_{j= -f}^{f} \frac{1}{1 - xz^{j}} . 
\label{key} \end{equation}
To illustrate how eq.(\ref{key}) can be used to obtain the structure 
of the maximum spin state
$|F,F\rangle$, we consider the following examples. 

{\em Spin-1 Bosons} :  For $f=1$, eq.(\ref{key}) gives
\begin{eqnarray}
 &  G^{(f=1)}(x,y) & = \frac{1}{(1-x^2)(1-xy)}    \\
 &  & = \sum_{n_{2} \geq 0}\sum_{\ell_{1} \geq 0} x^{2n_{2} +
\ell_{1}} y^{\ell_{1}} . 
\label{spin1gen} \end{eqnarray}
Comparing with eq.(\ref{gen}), we have 
\begin{equation}
M_{N}(S) = \sum_{n_{2} \geq 0}\sum_{\ell_{1} \geq 0} \Delta
\left( 2n_{2} + \ell_{1} -N \right) \Delta \left( \ell_{1} - F \right) . 
\label{xx} \end{equation}
Eq.(\ref{xx}) shows that $M_{N}(S)$ is the number of the solutions
of the equations 
\begin{equation} 
2\cdot n_{2} + 1\cdot \ell_{1} = N, \,\,\,\,\,\, 1\cdot \ell_{1} = F. 
\label{spin1con} \end{equation}
It is clear that eq.(\ref{spin1con}) has a unique solution 
$n_{0}=(N-F)/2, \ell_{1} =F$. Hence $M^{(f=1)}_{N}(F)=1$, i.e. 
there is only one maximum spin state $|F, F_{z}=F\rangle$. 
Next, we recall that the exponent
of $x$ and $y$ are associated with particle number and spin respectively.
Eq.(\ref{spin1con}) shows that the system consists of $\ell_{1}$ 
magnetic structural units which are spin-1 Bosons ($a_{1}$), and 
$n_{2}$ singlet pairs of Bosons $\Theta_{2}$. A simple 
exercise shows that $\Theta_{2} = 
(2a_{1}a_{-1} - a_{0}^2)$. The (un-normalized) many-body
state $|F, F_{z}=F\rangle$  is then given by 
$|F, F\rangle = a^{\dagger F}_{1}\Theta^{\dagger (N-F)/2}|{\rm vac}
\rangle$, 
which is the results given in  ref.(\cite{Hoyip}).  (See also fig.a). 

{\em Spin-2 Bosons} :  For $f=2$, eq.(\ref{key}) gives
\begin{equation}
G^{(f=2)}(x,y)  = 
\frac{1 + x^3 y^3}{(1-x^2)(1-x^3)(1-x y^2)(1- x^2 y^2) }. 
\label{spin2gen} \end{equation}
\begin{equation}
= \sum \sum_{m_{3}=0,1} x^{2s_{2}+3s_{3}+ m_{1} + 2m_{2} + 3m_{3}} 
y^{2m_{1} + 2m_{2} + 3m_{3}} ,
\end{equation}
where the first sum is over the non-negative integers set
$\{ s_{2}, s_{3}, m_{1}, m_{2} \}$. 
We then see that $M_{N}(F)$ is given
by the number of solutions to the equations 
\begin{equation}
2s_{2}+3s_{3}+m_{1} + 2m_{2} + 3m_{3} = N  , 
\label{spin2con1} \end{equation}
\begin{equation}
2m_{1} + 2m_{2} + 3m_{3}= F . 
\label{spin2con2} \end{equation}
A solution of eqs.(\ref{spin2con1}) and (\ref{spin2con2})
describes a state consisting $s_{2}$ two-particle singlets
$\Theta_2$,  $s_{3}$ three-particle singlets $\Theta_{3}$,
$m_{1}$ spin-2 Bosons ($a_{2}^{\dagger}$), and $m_{2}$
two-particle spin-2 state $|2,2\rangle$ (denoted as $\Gamma_{2}$). 
Since $m_{3}=0$ and 1, the 
system may or may not contain a three-particle spin-3 state $|3,3\rangle$ 
(dented as $\Gamma_{3}$) depending on whether $F$ is odd or even. 
It is straightforward to work out the expressions of these states, 
which are $\Theta_2 = a_{2}a_{-2} - a_{1}a_{-1} + \frac{1}{2}a_{0}^2$,
$\Theta_{3} = a_{0}^{}(2a_{2}^{}a_{-2}^{} + a_{1}^{}a_{-1}^{}-
\frac{1}{3}a^{2}_{0}) - \sqrt{\frac{3}{2}}(a^{2}_{1}a^{}_{-2} +
a^{}_{2}a^2_{-1})$,
$\Gamma_{2} = a^{}_{2}a^{}_{0} -  \frac{\sqrt{6}}{4}a^{2}_{1}$, and
$\Gamma_{3} = 2a^{2}_{2}a^{}_{-1} - \sqrt{6} a^{}_{2}a^{}_{1}a^{}_{0} +
a^{3}_{1}$.  The general structure of the state 
$|F, F\rangle$ is then $|F,F\rangle = \sum A(\{ s_{i}\}, \{ m_{i} \}) \left(
a^{\dagger m_{1}}_{2} \Gamma_{2}^{\dagger m_{2}} 
\Gamma_{3}^{\dagger m_{3}}\right) \Theta^{\dagger s_{2}}_{2} 
 \Theta^{\dagger s_{3}}_{3}  |{\rm vac}\rangle $. 

The condition $m_{3}=0,1$ is the additional constraint ${\cal L}_{\alpha}$
mentioned in section {\em (I.1)}. If $\Gamma_{3}$ was a ``free" unit that 
could appear as many as times as possible, the numerator of eq.(\ref{spin2gen})
would  be (instead of $1+x^3 y^3$) an infinite series 
$\sum_{m_{3}=0}^{\infty} (x^3 y^3)^{m_{3}}$, which will turn into 
a factor $(1-x^3y^3)^{-1}$ 
like other ``free" building units ($\Theta_{2},
\Theta_{3}, a_{2}$, and $\Gamma_{2}$) in the denominators in 
eq.(\ref{spin2gen}). The fact that the series of $x^3 y^3$ terminates at the
first order means that a pair of three-particle singlets can be expressed 
in terms of all other ``free" excitations ($\Theta_{2},
\Theta_{3}, a_{2}$, and $\Gamma_{2}$) and therefore has already been accounted
for in the generating function. Indeed, when examining $\Gamma_3$ (because of
the prediction of the generating function), one finds $\Gamma^2_3 =
-\frac{16\sqrt{6}}{9}\Gamma_2^3 + \frac{8\sqrt{2}}{3}a^2_2 \Gamma_2 \Theta_2 -
4\sqrt{\frac{2}{3}} a^3_2 \Theta_3$\cite{equivalent}. 
Note, however, that $\Gamma_{3}$ appears at most once. It therefore has no
thermodynamic significance. This means that one can obtain the relevant 
thermodynamic structure by taking any term in the 
numerator of eq.(\ref{spin2gen}). (See also fig.b). 

{\em (II). Bose gas with half integer spins} : 
When $f$ is a half integer, it
is useful to consider the generating function  
\begin{equation}
G(x,y) = \sum_{N\geq 0} \sum_{F\geq 0} M_{N}(F) x^{N}y^{2F}.
\label{halfgen} \end{equation}
Proceeding as the integer case, 
the function $W(x,z)$ in eq.(\ref{Wfinal})  becomes $W(x,z) = (1-z^{-2})
\prod_{j= -f}^{f}(1-xz^{2j})^{-1}$. Since $f$ is a half-integer, $W$
consists of even and odd powers of $z$. Using the
previous method to project out all the negative powers in $y$, we
have
\begin{equation}
G(x,y) = \int_{C} \frac{ {\rm d}z}{2\pi i} \frac{1-z^{-2}}{z-y}
\prod_{j= -f}^{f} \frac{1}{1 - xz^{2j}} . 
\label{halfcon} \end{equation}

{\em Spin 1/2 Bosons} : For $f=1/2$, we have 
\begin{equation}
G^{(f=1/2)}(x,y) = \frac{1}{1-xy}= \sum_{n\geq 0}
x^{n}(y^{2})^{n/2}
\end{equation}
Since the equation $n=N, n/2 =F$ has only one solution and forces
$F=N/2$.  This means that the total spin of a systems of  spin-1/2
Bosons is fixed by the particle number  $N$ to be $F=N/2$, and 
$|F, F=N/2 \rangle = a^{\dagger N}_{1/2}|{\rm
vac}\rangle$. The system can be referred to as a
``statistical ferromagnet" since the ferromagnetism is forced by
statistics. 

{\em Spin 3/2 Bosons} : For $f=3/2$, we have 
\begin{equation}
G^{(f=3/2)}(x,y) = 
\frac{1 + x^3 y^3}{(1-x^4)(1-x y^3)(1- x^2 y^2) }
\end{equation}
\begin{equation}
= \sum \sum_{m_{3}=0,1} x^{4s_{4}+ m_{1} + 2m_{2} + 3m_{3}}  
(y^2)^{\frac{3}{2}m_{1} + m_{2} + \frac{3}{2}m_{3}} , 
\end{equation}
where the first sum is over non-negative integers $s_{4}, m_{1}, m_{2}$. 
The number of spin state $|F, F\rangle$ is given by the number of the 
solution of
\begin{equation}
4 s_{4} + m_{1} + 2m_{1} + 3m_{3} = N, \,\,\,\,\,\, 
\frac{3}{2}m_{1} + m_{2} + \frac{3}{2}m_{3} = F, 
\end{equation}
which describes a state 
consisting of  $s_{4}$ four-particle singlets $\Theta_{4}$,  
$m_{1}$ spin-$3/2$ Bosons (i.e. $a_{3/2}$), and $m_{2}$
spin-1 pairs $|1,1\rangle$ made up of two spin-3/2 particles 
(denoted as $\Gamma_{2}$). Since $m_{3}=0$ and 1, the system 
may also contain a spin-3/2 
three-particle state $|\frac{3}{2},
\frac{3}{2}\rangle$, (denoted as $\Gamma_{3/2}$) which appears at most once. 
Thus, we have $|F,F\rangle = \sum A(\{ s_{i}\}, \{m_{i} \}) \left(
a^{\dagger m_{1}}_{3/2} \Gamma_{2}^{\dagger m_{2}} 
\Gamma_{3}^{\dagger m_{3}}\right) \Theta^{\dagger s_{4}}_{4} 
 |{\rm vac}\rangle$.  (See also fig.c). 

{\em (III) Spin-3 Bosons} : The case of $f=3$ begins to illustrate the full
complexity of the Bosons with higher spin. It is sufficiently intricate so we
discuss it last.  When $f=3$, eq.(\ref{con}) gives
\begin{equation}
G^{(f=3)}(x,y) = \frac{[1 + x^{15} + C(x,y)]D(x,y)}
{(1-x^2)(1-x^4)(1-x^6)(1-x^{10})}
\label{spin3gen} \end{equation}
\begin{equation}
D(x,y) = \frac{1}{(1-xy^3)(1-x^2y^2)(1-x^2y^4)}
\label{Ddef} \end{equation}
The term $C(x,y)$ is a polynomial with about fifty terms of the form 
$x^a y^b$ with ($a,b>0$). Since $b>0$, these terms represent magnetic 
structures. From eq.(\ref{gen}), we see that the structure of the total singlet 
state $|F=0, F_{z}=0\rangle$ is given by $G(x,y=0)$. Extracting 
$G(x,y=0)$ from eq.(\ref{spin2gen}) (i.e. setting $C=0$ and $D=1$), we see
that the singlet state is a linear combinations of singlets consisting of 
two, four, six, and ten particles, denoted as $\Theta_{2}, \Theta_{4}, 
\Theta_{6}, \Theta_{10}$ respectively. From our discussions for the spin-2 case,
we see that all singlet except that made up of 15 particles ($\Theta_{15}$) 
can be expressed as products and sums of the ``free" singlet set 
$\{ \Theta_{2}, \Theta_{4}, \Theta_{6}, \Theta_{10} \}$. However, two 
15-particle singlets (i.e. $\Theta_{15}^2$) is reducible to free singlet units.
(See also fig.d). 

As before, the elementary magnetic units $\{ \Gamma_{i} \}$ 
are given by the denominator of $D$. They are single particle spin-3 Bosons 
$(a_{3})$, two-particle  spin-2 pairs ($|2,2\rangle$), 
and two-particle spin-4 pairs ($|4,4\rangle$). 
The major difference between the $f=3$ and previous examples, however, 
is the appearance of large number of terms in the numerator of 
the generating function (i.e. $C$), and the fact that about half of 
these terms have 
negative signs, which means disappearance rather than appearance of 
a configuration. The origin of the 
negative terms is due to the fact that a product of two or more {\em different} 
magnetic units $\Gamma_{i}$ and $\Gamma_{j}$ can be expressed in terms of 
other magnetic and non-magnetic units. These are the ``interaction" constraints 
${\cal L}_{\alpha}$ we mentioned in section {\em (I.1)}.   
 Note that in the case of $f=2$, the interaction constraints comes from the 
reducibility of a single type of structure, 
i.e. $\Gamma^2_{3}$ is reducible into 
other free units. As a result, all terms in the numerator of $G^{(f=2)}$ are 
positive because one simply enumerates the multiplicity of $\Gamma_3$ until it
becomes reducible. If, however, the interaction constraints involve the
reducibility of the products of two or more {\em different} magnetic operators,
as well as ``scattering" such as $\Gamma_{i}\Gamma_{j} \rightarrow
\Gamma_{j}\Gamma_{k} + $etc, then the counting process can be not be simply a 
termination of the multiplicity of a particular pattern. We shall not analyse 
the interaction constraint for the $f=3$ case here because it is very involved. 
Despite this complexity, it is clear from the generating function what the
elementary magnetic building units are. 

In summary, we have illustrated the method to uncover the elementary 
building units of the angular momentum eigenstates of a spin-carrying 
Bose gas. The construction of these units is crucial for energy
studies\cite{HoYin}. 
Yet even without such studies, the present method has illustrated
the complex structure of the ground state of these Bose gases as a function of 
magnetization.  The fact that the number of
independent singlet units proliferates as $f$ increases also means that 
the system becomes more fragmented, since spin fluctuations (which is 
already huge in the spin-1 case in the low field limit\cite{Hoyip}) 
will increase as the number of different singlets increases. 

This work is supported by the NASA Grant NAG8-1441, the NSF Grants
DMR-9705295 and  DMR-9807284, and a Fellowship from John Simon 
Guggenheim Memorial Foundation.

\noindent Figure Captions : Figures a,b, and c are schematic representations of
the basis of the angular momentum state $|F, F\rangle$ for spin $f=1,2$, and 
3/2 respectively. Enclosed units with and without arrows represent magnetic and
singlet units respectively. The number of dots indicates the number of
particles in the unit. For example, in the spin-2 case (fig.b), the state
consists of 2-particle and 3-particle singlets (represented as 
arrow-free ellipses and triangles containing two and three dots resp.), 
and a 2-particle spin-2 pair 
(represented as an ellipse with two dots and an arrow).  
The 3-particle spin-3
unit is represented as a triangle containing 3 dots and an arrow.
The dashed circles in the interior are drawn to help to visualize the 
singlet and the magnetic units. They are not meant to imply the existence of a
singlet core. 
Figure d is a schematic representation of the singlet structure of a spin-3 
Bose gas, which consists of 2-, 4-, 6- , and 10-particle singlets, and a
``constraint" unit consisting of 15 particles, which is reducible to other 
existing singlet units when appears more than once. 


\begin{references}
\bibitem{equivalent} The reducibility of $\Gamma_{3}^2$ implies 
the choice of ``free" structural unit is not unique. Writing 
eq.(\ref{spin2gen}) as $G(x,y) = \frac{1+x^2 y^2 + x^4 y^4}
{(1-x^2)(1-x^3)(1-xy^2)(1-x^3y^3)}$ shows that the role of $\Gamma_2$ and 
$\Gamma_{3}$ can be interchanged. 
For higher spins, there can be many equivalent 
representations. One natural choice is to choose a representation where 
the ``free" units have minimum number of particles.
\end{references}
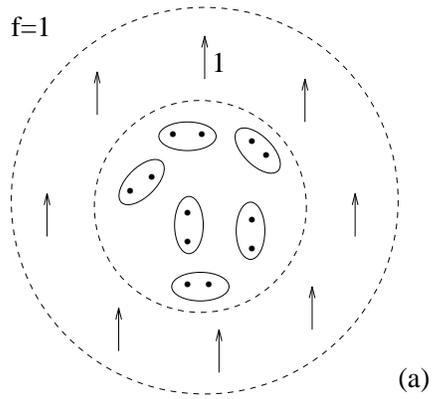
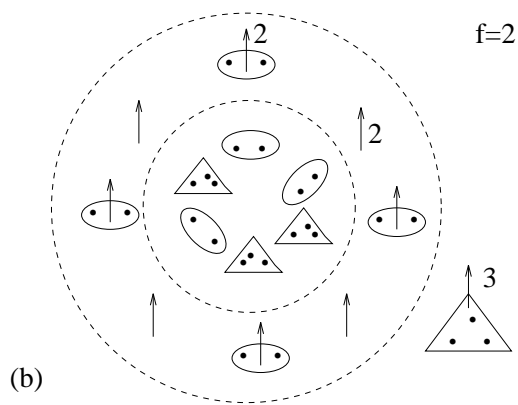
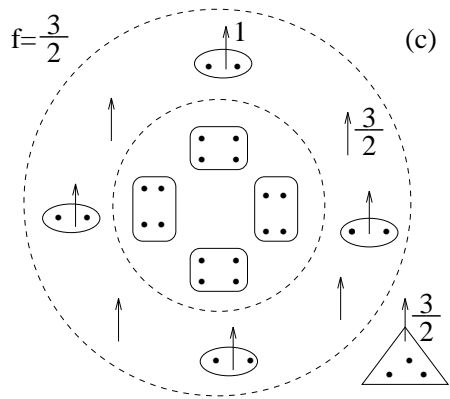
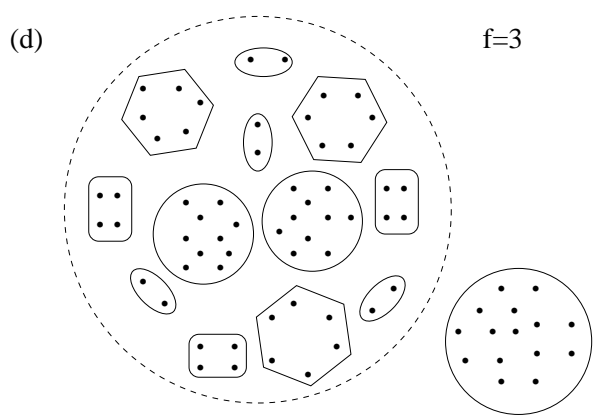
\end{document}